\documentclass[11pt]{article}

\usepackage[margin=1in]{geometry}
\usepackage[utf8]{inputenc}
\usepackage[T1]{fontenc}
\usepackage{microtype}
\usepackage[hidelinks]{hyperref}
\usepackage{graphicx}
\usepackage{booktabs}
\usepackage{amsmath}
\usepackage{amssymb}
\usepackage{xcolor}
\usepackage{array}
\usepackage{enumitem}
\usepackage{listings}
\usepackage{tikz}
\usetikzlibrary{arrows.meta,positioning,shapes.geometric,calc,fit}

\title{Everything Is a VisionBlock:\\
\large Conversational Authoring over Git-Versioned Content for Spatial Computing}
\author{%
  Zhaoming Yin \\
  \small Datumpont AI LLC, California, USA \\
  \small \texttt{zhaoming@datumpont.ai} \\
  \small \href{https://datumpont.ai}{datumpont.ai}
}
\date{\today}

\begin{document}
\maketitle

\begin{abstract}
Spatial applications compile their content into shipped binaries, so
every change costs a build-and-redeploy cycle. We present the
\emph{VisionBlock system}, which splits an application into an
\textbf{engine} --- a generic binary with a fixed set of capabilities
(render panels, volumes, and immersive scenes; fetch data; run
gestures) --- and \textbf{themes}: complete applications expressed as
trees of \textbf{VisionBlocks}, units of declarative content the engine
renders. Themes are data: creating, changing, or publishing one never
touches the binary. Authoring is a chat --- each turn produces a
VisionBlock's next version --- and versioning is plain git. The model is
five-dimensional: dimensions 1--3 are space (panel, volume, room);
dimension 4 is time (git history --- revert to roll back, branch to try
variants); dimension 5 is the principal (the per-user domain: the same
path resolves differently per person). The engine renders one point,
$(x, y, z, \mathit{version}, \mathit{principal})$. One consequence
follows per non-spatial axis: iteration collapses to chat turns and
reverts; ownership and permission are properties of content; and
together they make applications \emph{items} --- grantable, forkable,
sellable subtrees, an economy of apps inside one binary.

We report on a production implementation evaluated on Apple Vision Pro
(xrOS 26) across eight application stress cases spanning finance, document
viewing, 3D garment try-on, photorealistic 3D city streaming, voting, voxel
gaming, and cinema. As the engine's capability substrate matures, marginal
engine code drops asymptotically along a steep reuse curve ($191 \to 224 \to 7 \to 87 \to 0$ lines),
culminating in an interactive 364-voxel physics-enabled sandbox that required
zero new engine code. We further analyze systems findings unique to spatial
hardware: arm64e runtime concurrency discipline enforcing strict task-group
draining to avert allocator traps, and zero-UI immersion patterns replacing
floating cards with ambient spatial geometry.
\end{abstract}

\section{The idea}

An application splits into an \textbf{engine} --- a generic binary with a fixed set of capabilities (render panels, volumes, and immersive scenes; fetch data; run gestures) --- and \textbf{themes}: complete applications expressed as trees of \textbf{VisionBlocks}, units of declarative content the engine renders. Themes are data: creating, changing, or publishing one never touches the binary. Authoring is a chat --- each turn produces a VisionBlock's next version --- and versioning is plain git (\S3).

The model is \textbf{five-dimensional}: dimensions 1--3 are space (panel, volume, room); dimension 4 is time (git history --- revert to roll back, branch to try variants); dimension 5 is the principal (the per-user domain: the same path resolves differently per person). The engine renders one point: \emph{(x, y, z, version, principal)}.

One consequence per non-spatial axis: iteration collapses to chat turns and reverts (axis 4); ownership and permission are properties of content (axis 5, \S4); and together they make applications \textbf{items} --- grantable, forkable, sellable subtrees, an economy of apps inside one binary (\S4.7). A blockchain explorer, a document reader, an immersive showroom all run on the same engine; none requires a deploy to change.

While the preliminary design postulated this boundary in theory, this paper reports on the \textbf{production implementation and empirical evaluation on Apple Vision Pro (xrOS 26)}. To test whether this declarative abstraction holds under real-world spatial workloads, we deployed an empirical stress-test across eight distinct application genres (\S10): cross-theme app marketplaces, blockchain explorers, document viewers, interactive 3D garment fitting, 3D photorealistic city streaming, democratic ballot voting, interactive voxel sandboxes, and immersive cinema environments.

Through an explicit \textbf{Convergence Ledger}, we track the marginal engine code required to support each successive genre. Our results confirm a steep reuse curve ($191 \to 224 \to 7 \to 87 \to 0$ lines): once foundational capability classes for declared queries, binary asset streaming, and write transactions are established, authoring entirely new interactive spatial domains drops to \textbf{zero new engine lines}. Furthermore, deploying on physical spatial hardware surfaced critical systems insights: runtime concurrency invariants on arm64e hardware requiring strict task-group draining to prevent allocator crashes, and human-computer interaction principles dictating a transition from 2D floating windows to zero-UI ambient immersion.

\section{VisionBlock taxonomy}

A \textbf{VisionBlock} is the unit of content. Each one is made of four things:

\begin{itemize}[itemsep=2pt]
\item a \textbf{spec} --- declarative data saying what it is and how it looks;
\item \textbf{children} --- the VisionBlocks it contains;
\item a \textbf{chat} --- the conversation that creates and revises it;
\item a \textbf{history} --- every version it has ever had, in git.
\end{itemize}

Containment has three levels: a \textbf{theme} (L1, the whole app) contains \textbf{spaces} (L2, anything that occupies a surface), which contain \textbf{components} (L3, their constituents). Navigation edges, data queries, and style tokens are VisionBlocks too. The structure is general across apps and domains; expressiveness is gated by the engine's capability classes --- \S8 stress-tests it across domains.

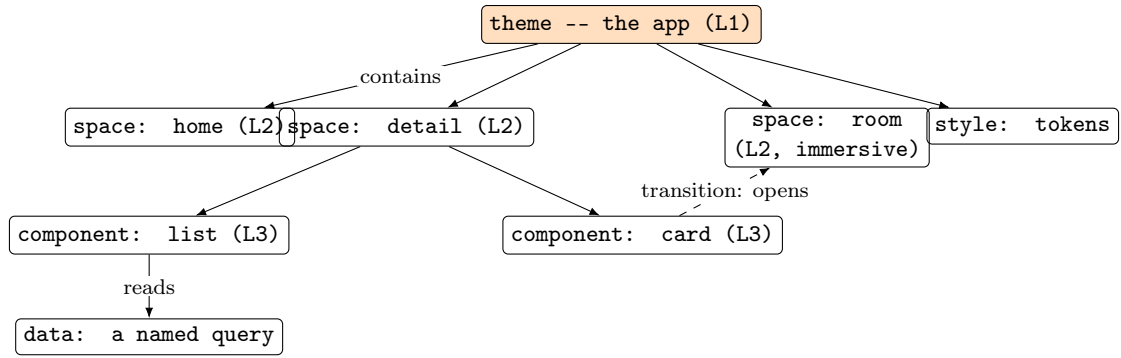
\begin{figure}[t]
\centering
\begin{tikzpicture}[
  n/.style={draw, rounded corners=2pt, font=\footnotesize\ttfamily,
            inner sep=3pt, align=center},
  e/.style={-{Latex[length=4pt]}},
  lab/.style={font=\scriptsize, fill=white, inner sep=1pt}]
\node[n, fill=orange!25] (T) {theme --- the app (L1)};
\node[n, below left=24pt and 70pt of T] (H) {space: home (L2)};
\node[n, below left=24pt and -20pt of T] (D) {space: detail (L2)};
\node[n, below right=24pt and -14pt of T] (R) {space: room\\(L2, immersive)};
\node[n, below right=24pt and 62pt of T] (S) {style: tokens};
\node[n, below left=26pt and -4pt of D] (L) {component: list (L3)};
\node[n, below right=26pt and -12pt of D] (C) {component: card (L3)};
\node[n, below=24pt of L] (Q) {data: a named query};
\draw[e] (T) -- node[lab]{contains} (H);
\draw[e] (T) -- (D);
\draw[e] (T) -- (R);
\draw[e] (T) -- (S);
\draw[e] (D) -- (L);
\draw[e] (D) -- (C);
\draw[e] (L) -- node[lab]{reads} (Q);
\draw[e, dashed] (C) -- node[lab]{transition: opens} (R);
\end{tikzpicture}
\caption{A theme is a tree of VisionBlocks: spaces contain components,
components read data VisionBlocks by name, and transitions are edges
stored as data.}
\label{fig:taxonomy}
\end{figure}

\subsection{L1 --- Theme VisionBlock (the app)}

The root. Owns:

\begin{itemize}[itemsep=2pt]
\item \textbf{Identity}: name, accent, app-wide copy.
\item \textbf{Data sources}: the external state providers this theme speaks to.
\item \textbf{Registry of L2 VisionBlocks}: which spaces exist, which one is home.
\item \textbf{Style VisionBlock reference}: palette, type scale, spacing tokens.
\end{itemize}

The engine boots by fetching a theme's resolved tree; a theme switcher lists available L1 VisionBlocks. The engine ships with no themes built in: every theme is user-created data.

\subsection{L2 --- Space VisionBlocks}

Anything that occupies a whole presentation surface:

\begin{itemize}[itemsep=2pt]
\item a 2D window
\item a bounded volumetric window
\item a full immersive space
\end{itemize}

\textbf{The L2 test is mutual exclusivity} (owner ruling, 2026-08-21): if a block claims the stage --- it cannot exist at the same time as another stage-claiming block --- it is an L2, whatever it is "about". Blocks that coexist alongside an L2 are its companions. The engine enforces this \textbf{structurally}, not by policing: there is ONE stage --- a single container, opened at launch and never closed --- and it renders one block at a time, so a second has nowhere to be, and a place is never lost because the window that holds it is never destroyed. \texttt{occupancy} says whether a block opens \emph{on} the stage or \emph{beside} it. The enforcement point is the engine's \textbf{stage} (\texttt{Sources/Stage.swift}): every surface the engine presents --- spec window, script volume, immersive room alike --- enters the stage on appear and a non-companion clears it of everyone else. The stage exists because visionOS cannot close what it cannot name: a value-based window group only dismisses \emph{with} the value it opened on, so the engine remembers which value is up. No surface dismisses another by hand, and every presentation goes through one doorway (\texttt{StageDoor}) rather than each call site holding its own opinion. \textbf{Every L2 has its own place}: a space's \texttt{place} (size, optional anchor) in its own data, which is the whole placement vocabulary visionOS accepts for a new window --- there is no absolute placement API, and no way to read where a user dragged a window. An L2 therefore opens in its declared place and never inherits another surface's: inheriting one (\texttt{.replacing}) inherits its \emph{transform}, and a volume's transform lay a reading window flat on a table. An immersive room is not a window and has no place to hold, so it takes the stage through a \textbf{keeper} window that carries the room's standard chrome. The engine names containers by \textbf{presentation mode only} --- one window group for \texttt{mode: window}, one for \texttt{mode: volume}, one immersive space for \texttt{mode: immersive}, each value-based on the space path --- so a theme with fifty volumes needs no engine build and no \texttt{ref}. Home is whatever \texttt{theme.json} declares; the engine carries no fallback space name. And \textbf{every L2 must carry at least one outward edge} --- back to main (the standard exit chrome provides this by construction), a link to another L2, or a door to its companions. A space with no way out is a bug by definition.

Spec: surface kind + size + layout of child L3 VisionBlocks + data bindings + \textbf{occupancy class} (which is the L2/companion distinction made machine-readable).

\textbf{Occupancy model} (the apartment rule): the user's focus is in exactly one room at a time, and the rules are data, not code ---

\begin{itemize}[itemsep=2pt]
\item \texttt{room} --- hard-exclusive full spaces (immersive). At most one exists; entering one exits the last. A man cannot be in two rooms at once.
\item \texttt{primary} --- the main surface of the moment. Exclusive within a mode: opening one replaces the current via its transition's handoff.
\item \texttt{companion} --- carried between rooms, never exclusive: side panels, the chat box, auxiliary lists. Furniture and hand-held items, not rooms.
\end{itemize}

Transitions declare handoff semantics (\texttt{replace | accompany | restore}), so navigation rules are authorable, versioned edges rather than code.

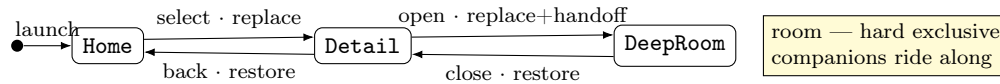
\begin{figure}[h]
\centering
\begin{tikzpicture}[
  st/.style={draw, rounded corners=3pt, font=\footnotesize\ttfamily, inner sep=4pt},
  e/.style={-{Latex[length=4pt]}, font=\scriptsize}]
\node[circle, fill, inner sep=1.6pt] (start) {};
\node[st, right=18pt of start] (H) {Home};
\node[st, right=64pt of H] (D) {Detail};
\node[st, right=76pt of D] (R) {DeepRoom};
\draw[e] (start) -- node[above]{launch} (H);
\draw[e] (H.10) -- node[above]{select $\cdot$ replace} (D.170);
\draw[e] (D.190) -- node[below]{back $\cdot$ restore} (H.-10);
\draw[e] (D.10) -- node[above]{open $\cdot$ replace+handoff} (R.170);
\draw[e] (R.190) -- node[below]{close $\cdot$ restore} (D.-10);
\node[right=10pt of R, font=\scriptsize, align=left, draw, fill=yellow!20,
      inner sep=3pt] {room --- hard exclusive\\companions ride along};
\end{tikzpicture}
\caption{Occupancy as data: one \texttt{room} or \texttt{primary} holds
focus at a time; transitions declare the handoff; companions accompany.}
\label{fig:occupancy}
\end{figure}

\subsection{L3 --- Component VisionBlocks}

The constituents of a space: stat bands, lists, cards, diagrams, 3D subscenes, control clusters. This level hosts the spec vocabulary the conversational authoring loop emits (cards, tiles, timelines, shapes) and grows (lists, grids, charts, tree/graph layouts).

\subsection{Transition VisionBlocks}

First-class edges, stored as data:

\noindent\begin{minipage}{\linewidth}
\begin{lstlisting}
{"type": "transition", "from": "detail-space.section-x",
 "gesture": "button", "label": "Open the deep view",
 "opens": "deep-room", "binding": {"id": "$selected.id"},
 "handoff": "replace"}
\end{lstlisting}
\end{minipage}

Every navigation an app would hand-code --- icons, deep-view buttons, back buttons, window handoffs --- becomes an authorable edge. Changing "what opens what" is a chat away, not a build.

\subsection{Other VisionBlock types this framework invites}

\begin{itemize}[itemsep=2pt]
\item \textbf{Data VisionBlocks} --- named, declarative queries against a source: URL template, extraction paths, refresh cadence, field typing. Components bind to data VisionBlocks by name --- \textbf{the single enabler of a second theme}: swap the data VisionBlocks, keep the spaces. v1: a generic JSON/REST driver; theme-specific native drivers may register as engine capabilities.
\item \textbf{Style VisionBlocks} --- design tokens referenced everywhere; a theme's accent flips in one place.
\item \textbf{Behavior VisionBlocks} --- parameterized choreography (spin, bob, arrival ceremonies) attachable to components.
\item \textbf{Voice VisionBlocks} --- the in-app explainer's persona per theme (a theme-level system prompt over the chat's per-topic context).
\end{itemize}

\section{Versioning and structure: native git, trees all the way down}

Versioning is a solved problem; the design delegates it to git rather than abstracting it. The delegation is unusually clean because both of the system's shapes are already git's shapes:

\begin{itemize}[itemsep=2pt]
\item \textbf{The containment structure is a tree} --- and a theme maps one-to-one onto a git tree: every VisionBlock is a file at its path, every parent a directory. The repository layout \emph{is} the L1/L2/L3 hierarchy.
\item \textbf{History is a tree too, not a chain.} Git's commit graph gives branching for free: fork an artifact to try two variants, keep both, merge the winner or abandon a branch. A linear history is just the degenerate single-branch case, not a designed constraint.
\end{itemize}

\noindent\begin{minipage}{\linewidth}
\begin{lstlisting}
themes/example.git                  one repository per theme
|-- theme.json                      L1 spec (identity, registry, sources)
|-- spaces/
|   |-- home/space.json             L2 spec
|   `-- detail/
|       |-- space.json              L2 spec
|       `-- components/
|           |-- list.json           L3 specs
|           `-- card.json
|-- transitions/*.json              the navigation edges
|-- data/*.json                     data VisionBlocks (queries)
|-- style/tokens.json
`-- chats/**                        transcripts, versioned beside the specs
\end{lstlisting}
\end{minipage}

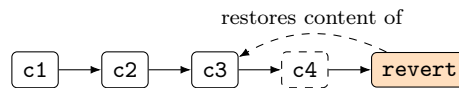
\begin{figure}[h]
\centering
\begin{tikzpicture}[node distance=10pt and 16pt,
  v/.style={draw, rounded corners=2pt, font=\footnotesize\ttfamily, inner sep=4pt},
  e/.style={-{Latex[length=4pt]}}]
\node[v] (c1) {c1};
\node[v, right=of c1] (c2) {c2};
\node[v, right=of c2] (c3) {c3};
\node[v, right=of c3, dashed] (c4) {c4};
\node[v, right=of c4, fill=orange!25] (r) {revert};
\draw[e] (c1) -- (c2);
\draw[e] (c2) -- (c3);
\draw[e] (c3) -- (c4);
\draw[e] (c4) -- (r);
\draw[e, dashed] (r) to[bend right=28] node[above, font=\scriptsize]{restores content of} (c3);
\end{tikzpicture}
\caption{Rollback in git terms: the revert commit restores c3's content
while c4 remains in history --- roll-forward is always possible, and the
log stays append-only.}
\label{fig:revert}
\end{figure}

\emph{Rollback in git terms: the revert commit restores c3's content while c4 remains in history -- roll-forward is always possible, and the log stays append-only.}

\subsection{The mapping, one-to-one}

\begin{center}\small
\begin{tabular}{@{}>{\raggedright\arraybackslash}p{0.30\linewidth}>{\raggedright\arraybackslash}p{0.65\linewidth}@{}}
\toprule
Git concept & VisionBlock concept \\
\midrule
repository & a theme --- one app, one repo \\
directory tree & the containment hierarchy (L1 / L2 / L3) \\
file (blob) & a VisionBlock's spec \\
commit & a chat turn --- author = principal, message = instruction + reply \\
log (per path) & a VisionBlock's history \\
revert & rollback --- itself a commit, so history stays append-only \\
branch & a variant draft --- merge the winner or abandon it \\
tag & a named release / pin \\
clone & forking a theme (the marketplace's \texttt{forkable} license) \\
diff & comparing any two versions \\
commit authorship & attribution, already per-version \\
\bottomrule
\end{tabular}
\end{center}

The blockchain resonance the earlier draft leaned on remains a pleasant pun --- append-only history, rollback as pointer movement --- but it is git's semantics, not a bespoke chain, and the design no longer abstracts it.

\subsection{Control plane stays outside the repository}

Git has no per-path access control, so the repository is never exposed raw: the server mediates every read and write through the \texttt{/vision} API, enforcing ACLs from the control-plane database. This preserves the rollback-trap invariant of \S4 with no extra machinery --- a git revert cannot touch a grant, because grants were never in the repository.

\noindent\begin{minipage}{\linewidth}
\begin{lstlisting}
-- control plane (Postgres): identity, entitlements, audit, quotas
block_acl   (theme, path, principal, role)
acl_audit   (theme, path, actor, change, at)
\end{lstlisting}
\end{minipage}

\subsection{API (unchanged semantics, git underneath)}

\noindent\begin{minipage}{\linewidth}
\begin{lstlisting}
GET  /vision/theme/{slug}          -> resolved tree at HEAD (one call to boot)
GET  /vision/block/{path}          -> spec at HEAD
GET  /vision/block/{path}/log      -> git log for that path
POST /vision/block/{path}/chat    -> one turn: reply + a commit
POST /vision/block/{path}/revert  -> restore an older blob (a new commit)
POST /vision/block                 -> create a file under a parent directory
\end{lstlisting}
\end{minipage}

Existing content stores migrate into the repository as VisionBlock files with their chats beside them;  prior read APIs remain as compatibility views over \texttt{git log}.

\section{Access control --- the business substrate}

Every VisionBlock carries (or inherits) an ACL. This section is load-bearing beyond security: \textbf{the business model runs on it.} Who may view, who may propose, who may decide, and who pays for what are all the same question, answered per VisionBlock.

\begin{figure}[h]
\centering
\begin{tikzpicture}[
  n/.style={draw, rounded corners=2pt, font=\footnotesize\ttfamily,
            inner sep=3pt, align=center},
  e/.style={-{Latex[length=4pt]}}]
\node[n] (L1) {L1 theme\\ACL: owner=all $\cdot$ public=view};
\node[n, below left=22pt and 58pt of L1] (A) {L2 home space\\inherits: public view};
\node[n, below=22pt of L1] (C) {L2 detail space\\inherits: public view};
\node[n, below right=22pt and 40pt of L1, fill=red!70!black, text=white]
  (B) {L2 fitting room (locked)\\explicit ACL: one user only};
\node[n, below left=20pt and -18pt of B, fill=red!70!black, text=white] (P) {photos\\private};
\node[n, below right=20pt and -16pt of B, fill=red!70!black, text=white] (AV) {avatar\\private};
\node[n, below=58pt of C, dashed] (X) {refused: cannot loosen to\\public below a private ancestor};
\draw[e] (L1) -- (A); \draw[e] (L1) -- (C); \draw[e] (L1) -- (B);
\draw[e] (B) -- (P); \draw[e] (B) -- (AV);
\draw[e, dashed] (B) -- (X);
\end{tikzpicture}
\caption{ACL inheritance: one grant at the root covers a theme; a private
subtree tightens at its own root; loosening beneath it is refused.}
\label{fig:acl}
\end{figure}
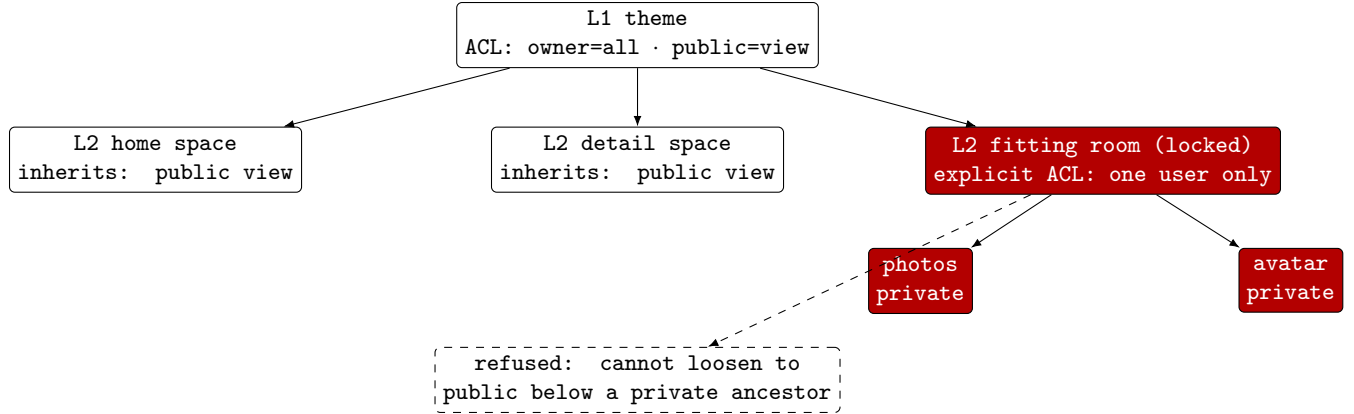

\subsection{Principals}

\begin{itemize}[itemsep=2pt]
\item \textbf{Owner} --- the account holder; the sole author until multi-principal grants ship.
\item \textbf{Users} --- real identities from an identity provider, with federated platform sign-ins (Apple, Google, Meta) riding on it as adapters. A device-level engine key is a transport gate, not an identity; the principal always comes from the identity layer.
\item \textbf{Viewers} --- anonymous read access, where granted.
\item \textbf{Agents} --- the LLM acts only \emph{on behalf of} a principal; every version records both the human principal and the model that generated it. No agent authority of its own.
\end{itemize}

\subsection{Rights, separated deliberately}

Per VisionBlock: \texttt{view} $\cdot$ \texttt{chat} (append versions) $\cdot$ \texttt{head} (rollback/forward) $\cdot$ \texttt{create-children} $\cdot$ \texttt{grant} (change the ACL itself). \texttt{head} is split from \texttt{chat} on purpose: letting someone propose versions is different from letting them decide which version is live --- contributor vs maintainer.

\subsection{Inheritance}

Effective ACL = nearest ancestor with an explicit ACL, walked up the containment tree. In practice one grant at the theme (L1) covers an entire app; a private subtree (dressing-room photos) sets an explicit stricter ACL at its root and everything beneath is private. Child ACLs may tighten freely; loosening below a private ancestor is refused --- no public VisionBlock inside a private subtree.

\textbf{Forks and grants.} Entitlements never travel with content: a clone carries the tree, never its ACLs. A fork mints a fresh ACL rooted at the forker, who owns the copy; and a subtree the principal cannot view cannot be cloned at all. Provenance stays in commit authorship; rights stay in the control plane.

\subsection{The rollback trap (why ACL is not spec)}

ACLs live in the control-plane database, \textbf{outside the repository}. If ACL were part of the spec, rolling a VisionBlock back would silently restore yesterday's looser permissions --- rollback as privilege escalation. So: specs roll back; ACLs never do. ACL changes append to an audit log instead, and \texttt{grant} right is required to touch them.

\noindent\begin{minipage}{\linewidth}
\begin{lstlisting}
block_acl   (block_id, principal, role)        -- role: owner|editor|contributor|viewer
acl_audit   (block_id, actor, change, at)      -- append-only, not rollbackable
\end{lstlisting}
\end{minipage}

Roles bundle rights: owner = all; editor = chat+head+children; contributor = chat only (proposes versions, a maintainer moves HEAD); viewer = view.

\subsection{Interactions with the rest of the design}

\begin{itemize}[itemsep=2pt]
\item \textbf{User-scoped VisionBlocks} (\S8.4) are just an ACL pattern: a subtree whose root grants owner-only to that user.
\item \textbf{Quota rides the principal}: chat turns, compute jobs and asset storage are metered per principal, which is also the abuse story for shared themes --- a viewer costs nothing, a contributor costs their own quota.
\item \textbf{Private assets} serve via short-lived signed URLs; the CDN never holds a public path to a private photo.
\item \textbf{Rollout}: the schema initially ships with a single implicit grant (owner: everything); real multi-principal grants arrive with user scoping, before any sharing feature.
\end{itemize}

\subsection{The business logic, expressed as ACL}

Every monetizable shape this product has discussed is a grant pattern --- no separate billing architecture needed at the model level:

\begin{center}\small
\begin{tabular}{@{}>{\raggedright\arraybackslash}p{0.30\linewidth}>{\raggedright\arraybackslash}p{0.65\linewidth}@{}}
\toprule
Business shape & ACL expression \\
\midrule
Free tier & \texttt{viewer} on public themes; zero quota consumed \\
Premium user & \texttt{contributor}/\texttt{editor} grants + a personal quota for chat, compute, assets \\
Paid theme (marketplace) & \texttt{view} grant on that theme's L1, sold; inheritance does the rest \\
Theme author royalties & authorship recorded per commit --- attribution is native to git \\
B2B analyst workspace & a private subtree: strict ACL at its root, client principals granted in; their traces, annotations and artifacts never leave it \\
Team seats & grants at L1 per principal; \texttt{contributor} proposes, a lead moves HEAD \\
Usage billing & quota metered per principal per right --- a viewer costs nothing, a builder pays their own way \\
\bottomrule
\end{tabular}
\end{center}

\subsection{Apps as items --- the theme marketplace}

The L1 VisionBlock is the sellable unit: \textbf{the engine is the shelf, themes are the products.} One engine binary per platform store; what's for sale inside them is the same data.

\begin{itemize}[itemsep=2pt]
\item \textbf{A product is a subtree + a listing.} Listing metadata (name, icon, preview captures, price, license) hangs off the L1 VisionBlock. Purchase mints a \texttt{view} grant; tiers can sell \texttt{contributor} (a theme you may extend) at a higher price. Inheritance delivers the whole app in one grant.
\item \textbf{Granularity is free.} Because any subtree is grantable, items smaller than apps fall out naturally: a single space (a signature room sold into someone else's theme), a component pack, a style pack. v1 sells L1 themes; the model already supports the rest.
\item \textbf{Updates ride git.} The author commits; buyers track the branch by default or pin a tag. A bad update is a revert, for the author or the buyer --- an affordance store-distributed binaries cannot offer.
\item \textbf{License is a field on the product}: \texttt{sealed} (view only), \texttt{forkable} (buyer may duplicate the tree as a base --- the clone operation from \S7, monetized), with attribution preserved in git history either way.
\item \textbf{Revocation and refunds are grant removals} --- auditable, instant, no binary involved.
\item \textbf{Platform payment rails are engine-local adapters}: StoreKit on Apple platforms, the corresponding billing APIs on Meta Horizon or Android XR. Mechanics are identical everywhere: platform receipt $\rightarrow$ server verification $\rightarrow$ grant minted. The grant store stays the single source of truth; every rail just feeds it --- and a theme bought on one platform is owned on all of them, because the grant, not the receipt, is the entitlement.
\item \textbf{The store is a theme.} The marketplace gallery is itself an L1 VisionBlock rendered by the engine --- browsing products, preview rooms, purchase transitions. The system sells things made of itself.
\end{itemize}

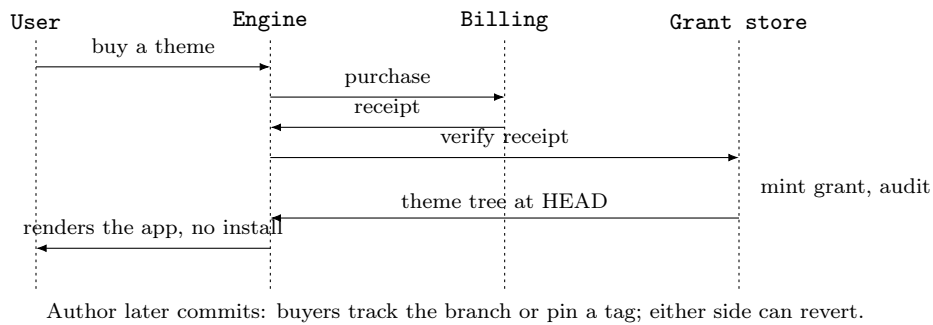
\begin{figure}[h]
\centering
\begin{tikzpicture}[font=\footnotesize\ttfamily,
  e/.style={-{Latex[length=4pt]}, font=\scriptsize}]
\node (U) at (0,0) {User};
\node (E) at (3.1,0) {Engine};
\node (SK) at (6.2,0) {Billing};
\node (S) at (9.3,0) {Grant store};
\foreach \x in {0,3.1,6.2,9.3} \draw[dash pattern=on 1pt off 1.5pt] (\x,-0.25) -- (\x,-3.55);
\draw[e] (0,-0.6) -- node[above]{buy a theme} (3.1,-0.6);
\draw[e] (3.1,-1.0) -- node[above]{purchase} (6.2,-1.0);
\draw[e] (6.2,-1.4) -- node[above]{receipt} (3.1,-1.4);
\draw[e] (3.1,-1.8) -- node[above]{verify receipt} (9.3,-1.8);
\node[font=\scriptsize, anchor=west] at (9.45,-2.2) {mint grant, audit};
\draw[e] (9.3,-2.6) -- node[above]{theme tree at HEAD} (3.1,-2.6);
\draw[e] (3.1,-3.0) -- node[above]{renders the app, no install} (0,-3.0);
\node[font=\scriptsize, align=left, anchor=west] at (0,-3.85)
  {Author later commits: buyers track the branch or pin a tag; either side can revert.};
\end{tikzpicture}
\caption{Purchase flow: whichever billing rail the platform requires
feeds one grant store; the grant, not the receipt, is the entitlement.}
\label{fig:purchase}
\end{figure}

The invariant that keeps this honest: \textbf{entitlements are grants, grants are auditable, and none of it rolls back with content.} Revenue logic never lives in specs.

\section{The engine (what the binary becomes)}

\textbf{One engine per platform, one content plane for all.} Because themes are pure data, porting the system is porting the engine: a visionOS engine, a Meta Quest engine, an Android XR engine --- even a desktop preview --- all implement the same capability contract (render panel / volume / immersive scene, data drivers, VisionBlock chrome, navigation) against the same repositories and grants. The contract includes the \textbf{scene API} --- the abstract interface (entities, meshes, gestures, attachments) that specs and script VisionBlocks target; each engine adapts it onto its platform's scene frameworks, the way browsers each implement one DOM. A theme authored on one platform runs on every platform whose engine speaks the contract; platform specifics (payment rails, script runtimes, sign-in) are engine-local adapters, never design constraints.

\textbf{The model layer is an abstraction, symmetric with the platform layer.} Chat turns target one provider interface --- context and transcript in, reply and next draft out. Concrete providers (aggregators, direct APIs, subscription-authenticated services, local models) are server-local adapters behind that interface; provider choice is data --- selectable per turn, per user, or per theme through voice VisionBlocks --- credentials are server configuration, and no provider name is architectural.

At render time the engine resolves the full tuple of \S1: spatial coordinates from specs and layout, \emph{version} from each VisionBlock's HEAD or pin, \emph{principal} from the identity layer --- one point of the five-dimensional space, every frame.

\begin{enumerate}[itemsep=2pt]
\item \textbf{Renderers} --- spec tree $\rightarrow$ panel, volume, immersive space, in the platform's native UI and scene frameworks. ArtifactPanelView/ArtifactVolumeView grow into the component vocabulary; two generic window groups ("space-panel", "space-volume") and one generic immersive space render \emph{any} L2 VisionBlock by id.
\item \textbf{Data drivers} --- domain-specific native drivers plus a generic declarative JSON/REST driver. A data VisionBlock compiles to: fetch, extract, type, cache, poll.
\item \textbf{Block chrome} --- every rendered VisionBlock can grow, on demand: a chat icon (chat), a version scrubber ($\blacktriangleleft$ N $\blacktriangleright$ --- rollback/forward, placed low so the user looks down onto controls, never past them), and an author-mode outline showing VisionBlock boundaries.
\item \textbf{Navigation interpreter} --- executes transition VisionBlocks (open/dismiss/ handoff patterns the app already uses, driven by data).
\item \textbf{Tolerant by rule} --- unknown kinds and node types render as nothing --- a load-bearing rule for the whole system: server vocabulary ships ahead of engine builds, old engines degrade gracefully.
\end{enumerate}

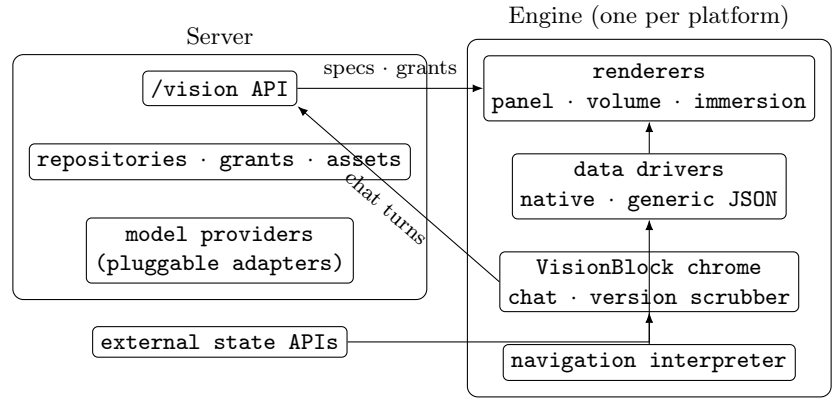
\begin{figure}[h]
\centering
\begin{tikzpicture}[
  n/.style={draw, rounded corners=2pt, font=\footnotesize\ttfamily,
            inner sep=3pt, align=center},
  grp/.style={draw, rounded corners=4pt, inner sep=6pt},
  e/.style={-{Latex[length=4pt]}, font=\scriptsize}]
\node[n] (API) {/vision API};
\node[n, below=14pt of API] (DB) {repositories $\cdot$ grants $\cdot$ assets};
\node[n, below=14pt of DB] (LLM) {model providers\\(pluggable adapters)};
\node[grp, fit=(API)(DB)(LLM), label={[font=\footnotesize]above:Server}] (SRV) {};
\node[n, right=70pt of API] (REN) {renderers\\panel $\cdot$ volume $\cdot$ immersion};
\node[n, below=12pt of REN] (DRV) {data drivers\\native $\cdot$ generic JSON};
\node[n, below=12pt of DRV] (CHR) {VisionBlock chrome\\chat $\cdot$ version scrubber};
\node[n, below=12pt of CHR] (NAV) {navigation interpreter};
\node[grp, fit=(REN)(DRV)(CHR)(NAV),
      label={[font=\footnotesize]above:Engine (one per platform)}] (ENG) {};
\node[n, below=16pt of LLM] (EXT) {external state APIs};
\draw[e] (API) -- node[above]{specs $\cdot$ grants} (REN);
\draw[e] (CHR.west) -- node[below, sloped]{chat turns} (API.south east);
\draw[e] (EXT.east) -| (DRV.south);
\draw[e] (DRV) -- (REN);
\draw[e] (NAV.north) -- (CHR.south);
\end{tikzpicture}
\caption{One content plane, one engine per platform: the capability
contract (renderers, drivers, chrome, navigation, and the scene API)
meets the server's repositories, grants, and pluggable model providers.}
\label{fig:engine}
\end{figure}

\section{Interaction model}

\begin{itemize}[itemsep=2pt]
\item \textbf{Author mode} is a toggle (owner-only for now, same app-key gate). Off: the app is just an app. On: VisionBlock boundaries glow faintly; every VisionBlock shows its chat icon and version scrubber.
\item \textbf{Chat scope rule}: a VisionBlock's chat mutates \emph{its own spec} and may create children. Rearranging or deleting children belongs to the parent's chat. Keeps blast radius contained and makes "which chat do I open" intuitive: talk to the thing you want to change.
\item \textbf{Rollback is not chat} --- it's the scrubber, instant and generation-free. Chat is for new versions; git history is for time travel.
\item The core journeys: empty VisionBlock $\rightarrow$ chat to create; existing block $\rightarrow$ render, refine by talking; swipe between sibling artifacts.
\end{itemize}

\section{The payoff: a second theme without a deploy}

\begin{enumerate}[itemsep=2pt]
\item Clone an existing theme's repository (a git clone).
\item In its chats: repoint the data VisionBlocks at the new domain's APIs; restyle via the style VisionBlock.
\item Every space renders the new domain's state immediately --- components were only ever bound to data VisionBlocks by name.
\item Where the new domain needs a new visual, chat creates component VisionBlocks in place.
\item Spaces that do not translate are simply absent from the new theme's registry --- themes list what they bear.
\end{enumerate}

Engine deploys are then reserved for: new node types in the vocabulary, new gesture kinds, new data-driver capabilities. Everything else is content. A concrete walk-through of this recipe --- Bitcoin to Ethereum --- is worked in \S8.2.

\section{Generality test and empirical evaluation}

Can the frame express arbitrary spatial apps? Verdict \emph{on paper}: \textbf{the structure generalizes; ambitious themes expose missing capability classes} --- the designed failure mode (capabilities ship as engine builds, everything else is data). The first case is the marketplace itself; every case after it is a theme that ships through that marketplace. Each worked in full, \textbf{on paper}:

The qualifier is load-bearing. Every case below was walked through as a thought experiment, and a thought experiment cannot fail the way a build can. \textbf{\S10.11 turns the argument into a measurement} --- each case built as a real theme on physical Apple Vision Pro hardware (xrOS 26), with what it costs the engine written down. Read the cases for the reasoning; read the ledger for whether it held.

\subsection{The universe --- the shelf every app is reached from}

The first proof of generality is the store: the theme marketplace of \S4.7 is not a special surface but an ordinary theme the engine renders.

\textbf{It is called the universe.} The owner's naming, 2026-08-28, and the name carries the design: a shelf is a list because paper is flat, and this is not paper. Every app is a star you walk among, and opening one is looking at it and tapping. "Marketplace" is what the thing sells; "universe" is what it \emph{is}, and the second is the part the engine has to render.

\noindent\begin{minipage}{\linewidth}
\begin{lstlisting}
L1 "Universe"
|- data:   catalog (GET /vision/themes -- every L1's own listing)
|- L2 "universe" (room)      -- one star per app, tapped to open       <- home
|- L2 "gallery"  (primary)   -- the same catalog as a flat index
|- L2 "preview"  (room)      -- walk a demo subtree of any listed theme
|- action: purchase          -- platform billing -> grant minted        [not built]
`- transitions: card -> preview (replace); preview -> purchase (action)
\end{lstlisting}
\end{minipage}

It exercises the platform end to end --- data VisionBlocks for the catalog, a room for previews, an \textbf{action VisionBlock} for purchase (the write capability of \S8.6's forms case, needed here first) --- and every case below ships \emph{through} it: each is a theme a creator could author, list, and sell without anyone deploying anything.

\textbf{And one thing this description missed.} The sentence above --- "not a special surface but an ordinary theme the engine renders" --- cannot be true while a theme has no way to address another theme. Previewing a listed theme, and going to one you have bought, are both \emph{cross-theme navigation}, and an address today names only a space. So the marketplace is not primarily a test of data, rooms and writes; it is the case that proves themes \textbf{compose}. See \S10.11 for the revised prediction and why it changed.

\subsection{A second blockchain explorer (Ethereum)}

The cheapest possible theme: the \S7 recipe instantiated, reusing every space of an existing theme by rebinding its data.

\begin{enumerate}[itemsep=2pt]
\item Clone the Bitcoin theme repository as a new theme (a git clone).
\item In its chats: "point data at Blockscout; blocks list from /api/v2/blocks; gas replaces fees; accent goes blue-violet" --- data + style VisionBlocks mutate.
\item The chain wall, byte layout and box system render ETH state immediately.
\item Where ETH needs a new visual (validators instead of pools), chat creates new component VisionBlocks in place.
\item Native VisionBlocks that don't translate (the BTC-specific fork tour) are simply absent from the ETH theme's registry.
\end{enumerate}

\begin{figure}[h]
\centering
\begin{tikzpicture}[
  n/.style={draw, rounded corners=2pt, font=\footnotesize\ttfamily,
            inner sep=3pt, align=center},
  grp/.style={draw, rounded corners=4pt, inner sep=6pt},
  e/.style={-{Latex[length=4pt]}, font=\scriptsize}]
\node[n, fill=orange!25] (ENG) {one engine binary\\(never redeployed for content)};
\node[n, below left=26pt and 30pt of ENG] (B1) {L1 first theme};
\node[n, below=12pt of B1] (BD) {data: chain A APIs};
\node[n, below=12pt of BD] (BS) {spaces: home $\cdot$ detail $\cdot$ rooms};
\node[grp, fit=(B1)(BD)(BS)] (BTC) {};
\node[n, below right=26pt and 30pt of ENG] (E1) {L1 forked theme};
\node[n, below=12pt of E1] (ED) {data: chain B APIs (rebound by chat)};
\node[n, below=12pt of ED] (ES) {same space specs, rebound};
\node[grp, fit=(E1)(ED)(ES)] (ETH) {};
\draw[e] (ENG) -- (B1);
\draw[e] (ENG) -- (E1);
\draw[e, dashed] (B1) -- node[above]{clone (a git operation)} (E1);
\end{tikzpicture}
\caption{The cheapest theme: clone an existing theme's repository and
rebind its data VisionBlocks --- every space renders the new domain
immediately.}
\label{fig:fork}
\end{figure}
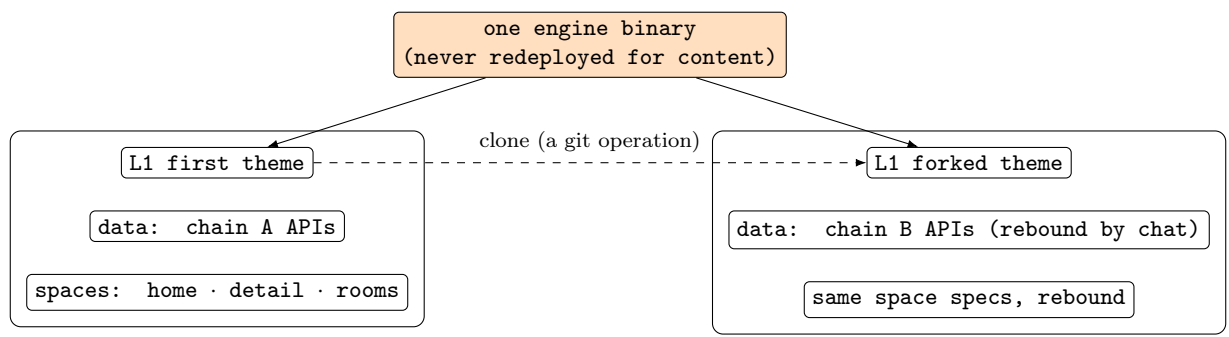

\subsection{A PDF reader occupying a space}

\noindent\begin{minipage}{\linewidth}
\begin{lstlisting}
L1 "Reader"
|- data:   library-index (list of documents)
|- L2 "shelf"    (primary window)  -- grid of documents, bound to library-index
|- L2 "reading"  (primary window)
|   |- L3 document-view   -> `document` node, ref: asset
|   `- L3 annotations     -- children created by chat/pinch, versioned
`- transitions: shelf.item -> reading (replace); reading -> shelf (restore)
\end{lstlisting}
\end{minipage}

\begin{itemize}[itemsep=2pt]
\item \textbf{New capability: asset VisionBlocks} --- binary content (the PDFs themselves) in a content-addressed store, referenced from specs. Base specs are JSON-only; this is the first wall every document-like theme hits.
\item New node type: \texttt{document} (PDFKit-backed), with page/scroll state.
\item Everything else --- shelf, navigation, annotations, versioning --- is the core system. \textbf{Cheapest of the stress cases.}
\end{itemize}

\subsection{The dressing room (user photos + a garment, rendered as a try-on)}

\noindent\begin{minipage}{\linewidth}
\begin{lstlisting}
L1 "Fitting Room"   (user-scoped theme)
|- assets: photos[] (PRIVATE, per-user)         <- user-scoped VisionBlocks
|- compute: avatar = photogrammetry(photos[])   <- compute VisionBlock, async
|- assets: garments[] (USDZ models)
|- L2 "mirror" (room)
|   |- L3 avatar-view  -> `model3d` node, ref: compute output
|   |- L3 garment-rack -- swipe garments (paged siblings)
|   `- L3 lighting     -- style VisionBlock
`- transitions: rack.item -> mirror re-dress (accompany)
\end{lstlisting}
\end{minipage}

\begin{itemize}[itemsep=2pt]
\item \textbf{New capability: user-scoped blocks.} By default a VisionBlock is global and owner-authored. Personal photos force per-user, private subtrees --- the authoring model grows an owner axis.
\item \textbf{New capability: compute VisionBlocks} --- declared async jobs: assets in $\rightarrow$ server-side pipeline (photogrammetry / avatar ML) $\rightarrow$ asset out, with status the chat surface can show ("building your avatar\ldots{}"). The VisionBlock system orchestrates; the ML itself is an external service behind the declaration, exactly as LLMs already are behind chat.
\item New node type: \texttt{model3d} (standard 3D model assets; platform scene frameworks render them nearly free).
\item \textbf{Hardest of the cases}, and every hard part is a capability class, not a taxonomy break: the tree above is ordinary blocks.
\end{itemize}

\subsection{The immersive property tour}

\noindent\begin{minipage}{\linewidth}
\begin{lstlisting}
L1 "Homes"
|- data:   listings (plain JSON driver -- a listings API)   ok already designed
|- L2 "map"   (primary volume)  -- listings on a globe/map
|- L2 "house" (room)
|   |- L3 environment -> `environment` node, ref: scanned-capture asset
|   |- L3 room-anchors[] -- subscene positions within the scan
|   `- L3 info-cards -- bound to listing fields, placed low in view
`- transitions: map.pin -> house (replace);
                room-anchor -> room-anchor (teleport)   <- new gesture kind
\end{lstlisting}
\end{minipage}

\begin{itemize}[itemsep=2pt]
\item Reuses: data driver (listings are just JSON), occupancy model (the house is literally a \texttt{room}), low-placed information cards.
\item New node type: \texttt{environment} (large scanned captures, needs asset streaming for size).
\item \textbf{Teleport} is one new gesture kind on the transition VisionBlock --- moving between rooms of the house is the apartment metaphor made literal, and it required no new block type at all.
\end{itemize}

\subsection{Forms and checkout flows}

Ordering, saving, submitting. Exposes: \textbf{action VisionBlocks} --- the data VisionBlocks are read-only queries; writes (POST with a payload schema, optimistic state, failure handling) are their own class. Form fields are ordinary L3 nodes; the submit edge is a transition whose target is an action.

\subsection{Media spaces}

A cinema screen or spatial-video wall: \texttt{media} node + streamed assets + playback state. Structurally trivial; the capability is codec/streaming support in the engine.

\subsection{Shared rooms}

Two people in one space (watching the same live scene, touring the same house). A whole new axis --- presence, shared block state, who mutates --- \textbf{deliberately post-v1}. The block model doesn't resist it (VisionBlocks are already server-side and versioned; concurrent viewers are natural), but authoring conflicts and presence are their own design doc.

\subsection{Games and simulations}

Apps whose essence is custom real-time logic. Expressible once \textbf{script VisionBlocks} land: the logic stored in the repository as code, LLM-generated in chat, executed by the embedded sandboxed runtime (an interpreter each platform's store rules permit).

\textbf{Shipped in P4.1} --- this section is kept as written because the prediction is the interesting part, but scripting is no longer on the road. The chain wall, the chain room, the fork ring, the pool globe, the miner graph and the merkle room are all script VisionBlocks: JavaScript in the repository, LLM-editable in chat, executed in a JavaScriptCore sandbox with no network and no filesystem. It was called "the last and largest capability layer" and it landed fourth.

\subsection{The pattern underneath}

These gaps re-derive the web platform layer by layer --- specs$\approx$HTML, style$\approx$CSS, data$\approx$fetch, actions$\approx$forms, assets$\approx$files, compute$\approx$workers, scripting$\approx$JS. Every stress case lands on a known layer instead of breaking the taxonomy --- the strongest available evidence the taxonomy is right.

\textbf{Capability roadmap, in demand order:} \textasciitilde{}\textasciitilde{}scripts\textasciitilde{}\textasciitilde{} (shipped P4.1) $\rightarrow$ \textbf{assets} $\rightarrow$ actions $\rightarrow$ user scoping $\rightarrow$ compute $\rightarrow$ environments $\rightarrow$ shared.

Assets is now the first wall, and the one most of \S10 is waiting behind: there is nowhere to put a PDF, a USDZ garment, a room scan or a video. Every block is JSON. A second \emph{data} explorer (\S10.2) is close to free; anything with a file in it needs this first.

That ordering is itself a prediction, and \S10.11 measures it: the roadmap is right if capabilities land in roughly this order because the cases demand them, and wrong if a case demands something the list does not contain.

\subsection{The convergence ledger: empirical measurement}
\label{sec:convergence-ledger}

Everything above was formulated as a structural prediction. To establish empirical rigor, we instantiated the owner's testing criterion:
\begin{quote}
\emph{"Instead of spending a week understanding the code, better to justify seeing how much diff 10.1 introduces, and then 10.2, 10.3 etc. The diff should diminish to 0 eventually."}
\end{quote}

We implemented each stress case as an actual, deployed theme on Apple Vision Pro (xrOS 26) and recorded the marginal engine source code required. If the capability taxonomy holds, the engine line cost falls toward zero as capability classes accumulate; if it does not, the engine is secretly an ad-hoc collection of domain-specific hacks.

\textbf{The critical metric is reuse.} A shrinking diff can be manufactured by introducing overly specific escape hatches. The honest signal of architectural convergence is \textbf{reuse}: whether case $N+1$ is served by the capability primitives introduced for case $N$. Table~\ref{tab:convergence-ledger} records the convergence ledger across all empirical stress runs.

\begin{table}[htbp]
\centering\small
\setlength{\tabcolsep}{3pt}
\begin{tabular}{@{}>{\raggedright\arraybackslash}p{0.18\linewidth}>{\raggedright\arraybackslash}p{0.13\linewidth}>{\raggedright\arraybackslash}p{0.34\linewidth}>{\raggedright\arraybackslash}p{0.31\linewidth}@{}}
\toprule
Case & Engine Lines & Capabilities Added & Capabilities Reused from Earlier \\
\midrule
\textbf{10.2 Ethereum explorer} \emph{(control)} & \textbf{55} (+12 debug) & Envelope path on query $\cdot$ ISO 8601 moment parsing $\cdot$ Hex prefix (\texttt{0x}) $\cdot$ Party relation (one/many) & --- (first empirical control case) \\
\midrule
\textbf{10.1 The universe} & \textbf{137} (+22 debug) & Block renders declared edges $\cdot$ \textbf{Qualified theme address} (\texttt{[theme:]space}) $\cdot$ List about nothing & \textbf{Envelope path} (catalog is an enveloped query) \\
\midrule
\textbf{10.2 again (app)} & \textbf{9} (+1 debug) & Wildcard path step matching arbitrary key (\texttt{*}) & \textbf{Envelope path} $\cdot$ \textbf{Hex prefix} $\cdot$ \textbf{ISO 8601} $\cdot$ Declared queries $\cdot$ Byte layout $\cdot$ Paging \\
\midrule
\textbf{10.1 again (labels)} & \textbf{15} & Declared label scale and plate styling & \textbf{Envelope path} $\cdot$ Declared queries $\cdot$ Catalog passthrough \\
\midrule
\textbf{10.3 PDF reader} & \textbf{191} (+1 debug) & Query returns raw bytes $\cdot$ Content-type declaration $\cdot$ URL argument $\cdot$ Surface wears returned image $\cdot$ Viewport embed sizing $\cdot$ Space \texttt{subject} fills query & \textbf{Declared queries} $\cdot$ \textbf{Host allowlist} $\cdot$ \textbf{Placeholder filling} $\cdot$ Script spaces $\cdot$ \texttt{events.json} $\cdot$ Transitions $\cdot$ Volume stage $\cdot$ Spec subject rule \\
\midrule
\textbf{10.4 Dressing room} & \textbf{224} (merged; 86 for \texttt{AssetCache}) & Surface wears \textbf{3D model} returned by query $\cdot$ Unified space \texttt{subject} $\cdot$ Binary typing $\cdot$ Theme-carried assets & \textbf{Bytes queries} $\cdot$ \textbf{Request shape} (\texttt{query}+\texttt{args}) $\cdot$ \textbf{Host allowlist} $\cdot$ \textbf{Placeholder filling} $\cdot$ Script spaces $\cdot$ Transitions $\cdot$ Volume stage \\
\midrule
\textbf{10.5 3D city streaming} & \textbf{7} (+44 hygiene) & Surface visible from \textbf{inside} (double-sided) $\cdot$ In-place subject mutation & \textbf{Bytes queries} $\cdot$ \textbf{\texttt{AssetCache}} $\cdot$ \textbf{Request shape} $\cdot$ \textbf{Host allowlist} $\cdot$ \textbf{Placeholder filling} $\cdot$ Immersive rooms $\cdot$ Texture projection \\
\midrule
\textbf{10.6 Forms (ballot)} & \textbf{87} (ballot: 0, write path: 87) & \textbf{Query write operation} (\texttt{POST}) $\cdot$ In-flight write tracking (\texttt{state.sending}) $\cdot$ Host form response & \textbf{Host allowlist} $\cdot$ \textbf{Placeholder filling} $\cdot$ \textbf{Request shape} $\cdot$ \textbf{Envelope path} $\cdot$ \textbf{Credentials} $\cdot$ \textbf{\texttt{AssetCache}} $\cdot$ \textbf{\texttt{ModelContext}} $\cdot$ Script spaces $\cdot$ HUD actions \\
\midrule
\textbf{10.9 Games (\texttt{craft} sandbox)} & \textbf{0} & None (complete capability saturation) & \textbf{Boxes with \texttt{touch} and \texttt{tap}} $\cdot$ \textbf{Panel rows} $\cdot$ \textbf{\texttt{hudAnchor}} $\cdot$ Script state \& \texttt{onTap} $\cdot$ Volume stage. 364 dynamic cubes rendered frame-to-frame \\
\midrule
\textbf{10.7 Media spaces (\texttt{cinema})} & \textbf{190} (+29 tests) & Declarative media screen (\texttt{media}) for streaming video $\cdot$ Universal clipboard paste verb (\texttt{paste}) & \textbf{Volume stage} $\cdot$ \textbf{Immersive rooms} $\cdot$ \textbf{Script spaces} $\cdot$ \textbf{Host allowlist} $\cdot$ \textbf{Request shape} $\cdot$ \textbf{HUD} $\cdot$ \textbf{Actions} $\cdot$ \textbf{\texttt{WKWebsiteDataStore}} \\
\bottomrule
\end{tabular}
\caption{The Convergence Ledger: marginal Swift engine source lines added versus capabilities reused across successive application stress cases deployed to Apple Vision Pro.}
\label{tab:convergence-ledger}
\end{table}

\subsection{Empirical stress case evaluations}

\subsubsection{10.2 Ethereum explorer (control): 55 engine lines}

The Ethereum explorer served as our experimental control. Structurally, Ethereum shares the same transaction graph, block sequence, and byte-level structure as Bitcoin. If the engine's core abstractions were truly domain-agnostic, rebinding the data source from Bitcoin's \texttt{mempool.space} API to an Ethereum EVM provider (Blockscout) should have cost approximately zero new engine lines.

The theme itself comprised 10 VisionBlocks, with its wall script modifying only \textbf{6 of 251 lines} from Bitcoin. Every node type, spatial spec renderer, byte map, transaction drill-in, and party list functioned untouched. 

However, running the live theme on Vision Pro revealed that the engine required \textbf{55 lines} of code across four distinct assumptions. None of these four was a fact about Ethereum; all four were implicit assumptions about how REST APIs format responses:
\begin{enumerate}[itemsep=2pt]
\item \textbf{Envelope paths} (24 lines): Blockscout wraps query responses in an envelope (\texttt{\{"items": [...]\}}), whereas the Bitcoin driver expected bare arrays. This envelope convention is ubiquitous across cursor-paged REST services (Stripe, GitHub).
\item \textbf{ISO 8601 timestamps} (15 lines): Blockscout returned timestamps as ISO 8601 strings (\texttt{2026-08-26T09:10:59Z}), whereas Bitcoin APIs provided Unix epoch integers.
\item \textbf{Hex prefix normalization} (6 lines): Ethereum addresses and hashes require \texttt{0x} prefixes, which the engine's byte-parsing previously rejected.
\item \textbf{One-vs-many entity relations} (9 lines): The engine had assumed sender and recipient entities were always arrays, whereas EVM transactions expose single sender/recipient pairs.
\end{enumerate}

These 55 lines were genuine generalizations that a single theme could never have surfaced. Furthermore, where Ethereum's block validation relies on Keccak-256 and RLP-encoded headers---primitives the engine lacked---the theme declared no cryptographic seal rather than fabricating a false one, demonstrating that declarative themes fail gracefully without engine corruption.

\subsubsection{10.1 The universe: cross-theme composition (137 engine lines)}

The universe is the spatial application launcher and marketplace (\S4.7). The preliminary design assumed the store was an ordinary theme requiring only transaction writes. When deployed, it exposed a fundamental architectural omission: \textbf{themes had no way to address other themes}.

If an address only names a space (\texttt{spaces/wall}), mounting a new theme leaves the stage stranded on an invalid path. Solving cross-theme navigation required \textbf{137 engine lines} centered on a single unifying abstraction: \textbf{qualified spatial addressing}:
\begin{equation}
\texttt{[theme:]space[?args]}
\end{equation}
When a theme declares an internal link, it writes an unqualified path (\texttt{spaces/wall}). Crucially, to prevent history corruption, \textbf{addresses are qualified at stage ingress}: as an edge is navigated, the engine stamps the currently mounted theme ID (\texttt{btc:spaces/wall}). Consequently, the system back-stack and history traversal remain globally unambiguous across theme boundaries without requiring theme authors to hardcode external namespaces. Catalog listings (\texttt{/vision/themes}) reused the envelope query path introduced by the Ethereum control.

\subsubsection{10.3 PDF reader: binary queries and viewports (191 engine lines)}

The document reader broke the engine's pure-JSON assumption. Documents, research papers, and books require binary byte streaming and dedicated spatial viewports. 

The reader added \textbf{191 engine lines}: introducing declared \texttt{bytes} queries, explicit content-type declarations, document viewport embedding via a native \texttt{document} node (backed by PDFKit), and dynamic query argument filling from the space's active \texttt{subject}. This case established the foundational binary transport layer, which subsequent 3D asset cases amortized immediately.

\subsubsection{10.4 Dressing room: 3D model streaming and asset caching (224 engine lines)}

The dressing room stress case streamed arbitrary 3D garments onto volumetric avatars. While declared \texttt{bytes} queries, placeholder filling, and host allowlists were reused cleanly from the PDF reader, rendering volumetric meshes required \textbf{224 engine lines}.

Initial implementations duplicated the network caching layer: a \texttt{ModelContext} was cloned from \texttt{TextureContext} (55 lines of identical cache, retry, and eviction logic differing only in the scene decoder). Recognizing this architectural duplication, we refactored both into a unified, content-addressed \textbf{\texttt{AssetCache}} (86 lines).

\textbf{Generative representations and spatial resolution.} In evaluating AI-generated 3D garment assets, we compared a human-specific research model (trained on rented GPUs) against commercial generative tooling. The research model emitted marching cubes with vertex coloring. On Apple Vision Pro, its color resolution was bounded by vertex density, producing blurry, muddy garments. Conversely, commercial tooling emitted UV-mapped polygonal meshes where 2.6 million texels were mapped independently of geometry. Re-baking 2.6M texels proved that geometric vertex resolution was never the visual bottleneck; spatial realism requires decoupled UV texture mapping.

\textbf{Stale asset cache invalidation.} Wearing the headset revealed a critical property of themes-as-data: editing an asset in-place in git did not refresh the headset display because \texttt{AssetCache} cached by URL arguments. We introduced explicit content-hash cache verification, validated by an automated continuous integration test asserting that an updated asset was refetched (\texttt{fetched 1x, wanted >= 2}).

\subsubsection{10.5 Photorealistic 3D city streaming: 7 engine lines}

The property tour and 3D geospatial city streamer placed real-world urban centers (Tokyo, Manhattan, Shenzhen) into volumetric tabletop boxes and immersive rooms. 

Remarkably, the engine diff was only \textbf{7 lines} (enabling double-sided polygon rendering for standing inside panoramic geometry, and in-place subject re-anchoring). The entire asset streaming, decoding, and caching pipeline was reused directly from \texttt{AssetCache} and declared queries.

The underlying streaming runtime connects to Google 3D Tiles, Project PLATEAU, and Hong Kong Lands Department open data. Operating on physical Vision Pro hardware, the streamer orchestrates \textbf{32 parallel tileset streams} and \textbf{48 concurrent glTF/Draco chunk decoders}, achieving sub-3-second first light. 

\textbf{Tabletop settled grounding.} Real-world geospatial coordinates place meshes on an Earth-centered ellipsoid, causing urban terrain to float or sink. The engine implements automatic settled grounding: the lowest mesh vertex within a 40-meter radius of the target coordinates is dynamically calculated and lifted to the room floor. This prevented street networks in Manhattan from sinking 200 meters below Times Square rooftops.

\subsubsection{10.6 Forms and midterm ballot: interactive write path (87 engine lines)}

Prior to this case, all queries were read-only GET requests. The midterm ballot stress case evaluated user interaction that mutates remote state: users gaze at 3D candidate models, tap to vote, and watch live tallies update.

The ballot itself cost \textbf{zero new engine lines}: candidate figures were rendered as standard 3D models via declared queries, tapping on models used existing event IDs, and dynamic tally bars were standard UI components.

Enabling state mutations cost \textbf{87 engine lines} in the engine's query substrate:
\begin{enumerate}[itemsep=2pt]
\item \textbf{Write queries} (\texttt{method: post}): Extending declared queries with POST semantics.
\item \textbf{Action verb dispatch}: Enabling scripts to invoke writes via \texttt{send:<query>}.
\item \textbf{In-flight state tracking} (\texttt{state.sending}): Exposing transaction status so themes prevent duplicate submissions on high-latency networks.
\end{enumerate}
Writes reused the existing host allowlist, credential validation, placeholder parameter substitution, and envelope unwrapping.

\subsubsection{10.9 Interactive voxel sandbox (\texttt{craft}): 0 engine lines}

The ultimate test of capability convergence was an interactive, physics-enabled 3D voxel sandbox (\texttt{craft}), inspired by Minecraft. The theme generates a procedural $16 \times 16$ terrain from a seed (incorporating water, sand, grass, dirt, stone, snow, and trees). 

The entire application cost \textbf{zero new engine lines} (PR \#328). The theme renders \textbf{364 dynamic interactive cubes} simultaneously on Apple Vision Pro. It reuses:
\begin{itemize}[itemsep=2pt]
\item \textbf{Direct spatial touch}: Reusing the fingertip collision slabs and tap events (\texttt{touch} and \texttt{tap}) originally developed for the virtual spatial keyboard.
\item \textbf{Spatial anchoring}: Reusing \texttt{hudAnchor} to float toolbars and block selectors in stable head-relative coordinates.
\item \textbf{Real-time scripting}: Sandboxed JavaScriptCore handling \texttt{onTap} raycasting to identify touched block faces, destroying touched voxels or placing newly selected materials.
\item \textbf{Delta state persistence}: World mutations are serialized as compact delta strings (e.g., \texttt{8,4,8=brick;9,4,8=}) and persisted to backend endpoint \texttt{/vision/keep/:slug/:key} via GET and POST, achieving instant save and restore without blocking modals.
\end{itemize}
Achieving a complete, interactive, real-time spatial voxel game with zero binary diff established that the engine's capability primitives had reached structural completeness for volumetric spatial apps.

\subsubsection{10.7 Media spaces (\texttt{cinema}): 190 engine lines}

The final stress case replicated Apple's native Cinema Environment: transforming an open room into a pitch-black, acoustically modeled movie theater playing high-resolution streaming video.

This case added \textbf{190 engine lines} (+29 unit tests), establishing generic media playback primitives without hardcoding service-specific logic:
\begin{itemize}[itemsep=2pt]
\item \textbf{Generic media declarations}: Added \texttt{script}, \texttt{css}, and \texttt{ambientGlow} to \texttt{DrawList.Media}, executed by an isolated \texttt{MediaWebView} in Swift. The engine acts strictly as a projector; service-specific styles and video players remain entirely within declarative theme scripts.
\item \textbf{Acoustic spatial geometry}: Theater seating is rendered across three distinct viewing tiers:
  \begin{itemize}
  \item \emph{Front Row}: 8.2\,m distance, low angle looking up, zero visual obstruction.
  \item \emph{Middle Row}: 12.5\,m THX reference distance, eye-level centered.
  \item \emph{Balcony / Back Row}: 17.0\,m elevated perspective looking down across the auditorium pit.
  \end{itemize}
  Ceiling slats and floor acoustic ribs terminate 2.5\,m before the virtual screen plane ($z = -12.5$\,m), leaving a clean architectural void around the video image.
\item \textbf{Persistent web session state}: Integrating \texttt{WKWebsiteDataStore.default()} to maintain authenticated user sessions across playback launches.
\item \textbf{Spatial alignment verbs}: Added \texttt{recenter} and \texttt{align} verbs to align the virtual cinema auditorium squarely with the viewer's head yaw in physical space.
\end{itemize}

\subsection{Empirical systems and HCI insights}

\subsubsection{The reuse curve: asymptotic convergence}

\begin{figure}[t]
\centering
\begin{tikzpicture}[scale=0.85]
\draw[->, thick] (0,0) -- (9,0) node[right, font=\footnotesize] {Stress Case Chronology};
\draw[->, thick] (0,0) -- (0,5.5) node[above, font=\footnotesize] {New Engine Lines};

\draw[help lines, color=gray!30, dashed] (0,0) grid (8.5,5);

\node[below, font=\scriptsize] at (1,0) {10.3 (PDF)};
\node[below, font=\scriptsize] at (2.5,0) {10.4 (Dress)};
\node[below, font=\scriptsize] at (4,0) {10.5 (City)};
\node[below, font=\scriptsize] at (5.5,0) {10.6 (Form)};
\node[below, font=\scriptsize] at (7,0) {10.9 (Craft)};

\node[left, font=\scriptsize] at (0,0) {0};
\node[left, font=\scriptsize] at (0,1.74) {87};
\node[left, font=\scriptsize] at (0,3.82) {191};
\node[left, font=\scriptsize] at (0,4.48) {224};

\coordinate (P1) at (1, 3.82);
\coordinate (P2) at (2.5, 4.48);
\coordinate (P3) at (4, 0.14);
\coordinate (P4) at (5.5, 1.74);
\coordinate (P5) at (7, 0);

\draw[thick, color=blue!80!black, mark=*, mark options={fill=blue!80!black}] 
  plot coordinates {(P1) (P2) (P3) (P4) (P5)};

\node[above right, font=\scriptsize, color=blue!80!black] at (P1) {191};
\node[above, font=\scriptsize, color=blue!80!black] at (P2) {224};
\node[above, font=\scriptsize, color=blue!80!black] at (P3) {7};
\node[above right, font=\scriptsize, color=blue!80!black] at (P4) {87};
\node[above right, font=\scriptsize, color=blue!80!black] at (P5) {\textbf{0 lines}};

\draw[dashed, red!80!black, thick] (0.8,4.2) to[out=-20,in=175] (7.2,0.1);
\node[above right, font=\scriptsize, color=red!80!black] at (3.5,2.6) {Reuse Convergence};
\end{tikzpicture}
\caption{The Reuse Curve: Marginal Swift engine lines added across successive application genres. As fundamental capability classes (binary caching, model loading, write transactions) are absorbed into the engine substrate, the cost of authoring entirely novel spatial application domains drops asymptotically to zero.}
\label{fig:reuse-curve}
\end{figure}

The central thesis of the VisionBlock architecture is that spatial applications decompose into a finite, closed set of generic capabilities. Figure~\ref{fig:reuse-curve} illustrates the empirical trajectory of engine modifications across our stress cases.

Rather than remaining constant or growing with application complexity, the required engine diff exhibits an asymptotic decline:
\begin{equation}
191 \longrightarrow 224 \longrightarrow 7 \longrightarrow 87 \longrightarrow 0
\end{equation}
The initial peak at 10.3 and 10.4 represents the unavoidable investment in fundamental capability classes: binary asset streaming, content-type parsing, and unified 3D mesh caching. Once \texttt{AssetCache} and model decoders existed, subsequent complex visual applications amortized that substrate: photorealistic 3D city streaming required only 7 lines, and the 3D midterm voting ballot required zero lines for its entire visual presentation. 

By the time the voxel sandbox (\texttt{craft}) was implemented, all required primitives---direct fingertip collision, volume bounding, HUD positioning, real-time sandboxed JavaScript execution, and asynchronous delta persistence---already existed. The engine diff dropped to \textbf{zero}. The capability substrate had saturated.

\subsubsection{Physical hardware vs.\ simulator concurrency: arm64e stack discipline}
\label{sec:hardware-concurrency}

Deploying 3D tile streaming to physical Apple Vision Pro hardware uncovered a critical concurrency defect that was completely invisible on the macOS xrOS Simulator.

\textbf{The runtime crash.} During rapid user panning across 3D city tilesets, the app suffered sudden, non-deterministic crashes with \texttt{SIGABRT} in the Swift runtime:
\begin{lstlisting}
libswift_Concurrency.dylib`swift_task_dealloc + 124
libswift_Concurrency.dylib`swift::swift_Concurrency_fatalError
VisionBlock`TileStreamer.streamOnce() (TileStreamer.swift:640)
fatalError: "deallocating task local allocator out of order"
\end{lstlisting}

\textbf{Root cause analysis.} Swift Concurrency manages task execution memory using a slab-based task-local allocator requiring strict Last-In, First-Out (LIFO) stack deallocation. In our initial implementation of \texttt{TileStreamer.swift}, parallel tile chunk downloads were dispatched using \texttt{withTaskGroup}:
\begin{lstlisting}
// Defective implementation: early exit causes LIFO mismatch
await withTaskGroup(of: (URL, Result<Decoded?, any Error>).self) { group in
    for tile in toLoad { group.addTask { ... } }
    for await (u, result) in group {
        if Task.isCancelled {
            group.cancelAll()
            return // FATAL TRAP on arm64e hardware
        }
    }
}
\end{lstlisting}
The \texttt{for await ... in group} loop allocates a synthetic \texttt{AsyncIterator} frame on the task's stack slab. When \texttt{Task.isCancelled} was detected, executing \texttt{group.cancelAll()} followed by an immediate \texttt{return} exited the closure while child tasks were still executing asynchronously. When \texttt{withTaskGroup} attempted to deallocate its own stack frame, \texttt{swift\_\allowbreak task\_\allowbreak dealloc} detected an out-of-order pointer mismatch on the stack slab and trapped with \texttt{SIGABRT}.

\textbf{Why the simulator succeeded.} The macOS xrOS Simulator runs as an x86\_\allowbreak 64 or host arm64 process under macOS user-space allocator semantics, which did not assert on task-local stack slab ordering invariants. Physical Apple Vision Pro hardware runs Apple silicon arm64e under visionOS 26, which rigorously enforces task-local slab deallocation ordering at the microarchitectural level.

\textbf{The architectural resolution.} To guarantee clean LIFO deallocation, the task group must never be abandoned while child tasks remain in-flight. We replaced the \texttt{for await} loop with explicit child task draining:
\begin{lstlisting}
// Correct arm64e runtime discipline: explicit child draining
await withTaskGroup(of: (URL, Result<Decoded?, any Error>).self) { group in
    for tile in toLoad { group.addTask { ... } }
    while let (u, result) = await group.next() {
        if Task.isCancelled {
            group.cancelAll()
            while await group.next() != nil {} // Drain in-flight tasks
            return
        }
        // Process tile result...
    }
}
\end{lstlisting}
By draining \texttt{group.next()} until exhaustion, all child task execution frames unwind in exact reverse allocation order, guaranteeing that the task stack slab restores its original pointer before \texttt{withTaskGroup} returns.

\subsubsection{Zero-UI immersion in spatial computing}
\label{sec:zero-ui}

Spatial user interfaces frequently suffer from "window clutter": floating 2D panels and buttons anchored in front of 3D content, which break immersion, occlude objects, and cause visual fatigue. 

Through iterative in-headset evaluation of the Cinema and City themes on Vision Pro, we formulated three human-computer interaction principles for spatial immersion:

\begin{enumerate}[itemsep=3pt]
\item \textbf{Eliminating 2D inspectors for ambient geometry}: Early iterations of the cinema space featured floating inspector cards displaying movie metadata and settings. Users reported feeling like they were sitting in an office looking at a media widget. By eliminating cards and embedding visual cues directly into the room geometry---pitch-black auditorium ceilings during film playback (\texttt{lights: false}, \texttt{\#070709}) and dark matte stepped seating silhouettes---the interface recedes entirely, leaving only the monumental content.
\item \textbf{Transient glass capsule dock with auto-dismiss}: Rather than anchoring persistent chrome around the user, controls are entirely hidden by default (\texttt{hud: null}, \texttt{actions: []}). A single \texttt{SpatialTapGesture()} anywhere in space catches user pinches and summons a compact, low-slung horizontal glass capsule dock using \texttt{.glassBackgroundEffect} positioned at eye level below the content ($[0.0, 0.70, -2.0]$\,m). An idle background timer (\texttt{idleDismissTask}) automatically dismisses the dock after 4 seconds of inactivity, smoothly returning the environment to pure immersion.
\item \textbf{Hit-test disabling on empty RealityKit entities}: In visionOS, invisible floating UI containers can inadvertently intercept user eye gaze and hand gestures. When \texttt{drawList.hud} is empty, the engine explicitly sets \texttt{hud.isEnabled = false} on the RealityKit attachment entity. This completely eliminates hit-testing overhead and prevents invisible UI bounds from intercepting raycasts targeted at 3D scene elements.
\item \textbf{Preserving system palm affordances}: Suppressing system overlays via \texttt{.persistent\-System\-Overlays} inadvertently disabled visionOS's wrist-look gesture (the floating white dot enabling access to Control Center), trapping users in immersion. Restoring \texttt{.automatic} overlays preserved essential accessibility without compromising the spatial experience.
\end{enumerate}

\section{Related systems}

The design competes with none of the systems below; it sits downstream of all of them --- declarative frameworks below the engine, generation models behind the provider interface, standards below the scene API --- and what it claims as new is the combination.

\textbf{End-user authoring.} HyperCard~\cite{hypercard} made applications documents their users could author, and its stacks remain the closest ancestor of themes: complete applications as shippable content. The VisionBlock system inherits that ambition and replaces scripting with conversation. Contemporary LLM interface generators such as v0~\cite{v0} demonstrate that models can produce working interfaces from prompts; the design consumes that capability rather than competing with it --- any spec-emitting model can sit behind the provider interface of \S5. The difference is what happens next: their output enters a developer's build pipeline, while here the model's output is committed, versioned, and rendered live, and the transcript that produced it is the version history.

\textbf{Declarative UI.} React~\cite{react} and SwiftUI~\cite{visionos} established the view as a pure function of declarative state, but the declarations compile into the shipped binary. The engine/theme split moves them out of the binary into versioned data; the capability classes of \S5 play the role a framework's component set plays, with the difference that exercising them never requires a build.

\textbf{Versioned content.} Git-backed data systems such as Dolt~\cite{dolt} give a database branch, diff, and merge semantics; event sourcing~\cite{fowler2005} rebuilds state from an append-only log. The design takes the same position for spatial content but binds the unit of history to the unit of authorship: one chat turn, one commit (\S3). CRDTs~\cite{shapiro2011} merge concurrent edits without coordination; the design instead serializes each VisionBlock's history through its chat, and shared-room co-editing is the point where CRDT semantics would enter (\S10).

\textbf{Authorization.} Zanzibar~\cite{zanzibar2019} is the reference point for relationship-based access control at global scale. The \texttt{block\_\allowbreak acl} table of \S4 is a deliberately small relative --- rights attach to tree paths and inherit downward --- and the claim made for it is not mechanism but role: the grant table doubles as the commercial substrate, so a purchase, a subscription, and a B2B seat are the same row.

\textbf{Creation platforms.} Roblox~\cite{roblox} and Meta Horizon Worlds~\cite{horizonworlds} are the nearest commercial analogues: many user-created experiences, and in Roblox's case a creator economy, inside one binary. The differences are the authoring surface (an editor plus scripting for a creator class, versus conversation for anyone), the versioning model (opaque platform state, versus git with revert, branch, and fork), and the target (games and social venues, versus applications).

\textbf{Spatial standards.} OpenXR~\cite{openxr}, the WebXR Device API~\cite{webxr}, and OpenUSD~\cite{usd} standardize device access and scene interchange. The scene API of \S5 is a smaller, higher-level boundary --- capability classes rather than geometry --- and an engine could implement it over any of them; they are complements below the portable surface, not alternatives to it.

What the design claims as a combination, none of these systems offer together: conversational authoring whose transcript is the version history, git as the only versioning machinery, an ACL model that is also the business model, and a marketplace of applications inside one binary.

\section{Open problems}

\begin{itemize}[itemsep=2pt]
\item \textbf{The vocabulary is the moat and the bottleneck.} Every visual the specs can't express falls back to native code. Grow node types by demand, never speculatively.
\item \textbf{Full-draft mutation won't scale to L2 specs.} In the base design every chat turn emits the whole spec. Fine for artifacts; a space spec will need patch-style edits (JSON-merge emitted by the model) as spec sizes grow.
\item \textbf{Perf in immersion}: spec-rendered scenes need the same discipline the hand-built ones learned (persistent scenes rebuilt incrementally, batched meshes, entity budgets) built into the renderer once, not per scene.
\end{itemize}

\section*{Acknowledgments}
The system, and this paper, were built in continuous collaboration with an
AI assistant (Claude, Anthropic) operating the development toolchain under
the author's direction; per arXiv policy the assistant is not an author.
The design crystallized through in-headset corrections whose verbatim
history is preserved in the project's engineering journal.

\nocite{git,nakamoto2008,visionos,hypercard,fowler2005,openrouter,mempool,blockscout}
\bibliographystyle{plain}
\bibliography{references}

\begin{thebibliography}{10}

\bibitem{visionos}
{Apple Inc.}
\newblock {visionOS}: {SwiftUI}, {RealityKit}, and volumetric scenes.
\newblock Developer documentation, 2026.

\bibitem{blockscout}
{Blockscout}.
\newblock Open-source {EVM} block explorer and {API}, 2026.

\bibitem{git}
Scott Chacon and Ben Straub.
\newblock {\em Pro Git}.
\newblock Apress, 2nd edition, 2014.

\bibitem{dolt}
{DoltHub}.
\newblock Dolt: A version-controlled {SQL} database, 2026.

\bibitem{fowler2005}
Martin Fowler.
\newblock Event sourcing, 2005.

\bibitem{hypercard}
Danny Goodman.
\newblock {\em The Complete HyperCard Handbook}.
\newblock Bantam Books, 1987.

\bibitem{horizonworlds}
{Meta Platforms}.
\newblock Meta horizon worlds, 2026.

\bibitem{react}
{Meta Platforms}.
\newblock React: The library for web and native user interfaces, 2026.

\bibitem{nakamoto2008}
Satoshi Nakamoto.
\newblock Bitcoin: A peer-to-peer electronic cash system.
\newblock White paper, 2008.

\bibitem{openrouter}
{OpenRouter}.
\newblock A unified interface for {LLMs}, 2026.

\bibitem{zanzibar2019}
Ruoming Pang, Ramon Caceres, Mike Burrows, Zhifeng Chen, Pratik Dave, Nathan
  Germer, Alexander Golynski, Kevin Graney, Nina Kang, Lea Kissner, Jeffrey~L.
  Korn, Abhishek Parmar, Christina~D. Richards, and Mengzhi Wang.
\newblock Zanzibar: {Google}'s consistent, global authorization system.
\newblock In {\em USENIX Annual Technical Conference}, 2019.

\bibitem{usd}
{Pixar Animation Studios}.
\newblock {OpenUSD}: Universal scene description, 2026.

\bibitem{roblox}
{Roblox Corporation}.
\newblock Roblox creator hub, 2026.

\bibitem{shapiro2011}
Marc Shapiro, Nuno Pregui{\c{c}}a, Carlos Baquero, and Marek Zawirski.
\newblock Conflict-free replicated data types.
\newblock In {\em Stabilization, Safety, and Security of Distributed Systems
  (SSS)}, 2011.

\bibitem{openxr}
{The Khronos Group}.
\newblock {OpenXR}: Cross-platform {XR} standard, 2026.

\bibitem{mempool}
{The Mempool Open Source Project}.
\newblock mempool.space {REST} {API}, 2026.

\bibitem{v0}
{Vercel}.
\newblock v0: Generative user interface design, 2026.

\bibitem{webxr}
{W3C}.
\newblock {WebXR} device {API}.
\newblock W3C Candidate Recommendation, 2026.

\end{thebibliography}

\end{document}